# Observational Accuracy of Variable Stars, Novae and Supernovae from Naked Eye to General Relativistic Standard: a Balance over Thousand SGQ Observations Sent to AAVSO.

Costantino Sigismondi, ICRA and G. Ferraris Institute, Rome, Italy.
sigismondi@icra.it

The theory of General Relativity deals with very accurate measurements that show significant divergences from Newtonian predictions only with speeds near to the velocity of light. An introduction for educational purposes, based on naked eye photometry, deals with the radiation near collapsing star's shells like novae and supernovae. The theme of accuracy is drafted from entry level observations to the precision of professional data, often of public domain on the web.
Thousand observations of variable stars, included the type 1a SN2014J, the Nova Del 2013 and the Nova Cen 2013, sent to the AAVSO by the author, with SGQ code, during the period 1998-2015 are analyzed to increase the photometric accuracy, in the occasion of the International Year of Light 2015.

### Introduction: AAVSO and "citizen astronomy"

The American Association of Variable Stars Observers, AAVSO, has been founded in 1911, and celebrated its centennial year in 2011[1] with now more than 28 million observations stored in its online database at the website www.aavso.org
W. T. Olcott[2] founded AAVSO to monitor 100 irregular variable stars of variable stars coordinated observations, in the fall of 1911 there were 13 members including 3 women, now the association's members are thousands and spread all over the World.
Before AAVSO the study of variable stars was conducted principally by professional astronomers and the results were gathered in special publications like the book of Paul Guthnik[3] on the history of Mira Ceti observations and the volumes edited by F. W. A. Argelander,[4] the promoter of variable stars studies in the XIX century. AAVSO archive contains thousands of observations of XIX and XX century sent by the observers and gathered by the scholars of the beginning of XX century.

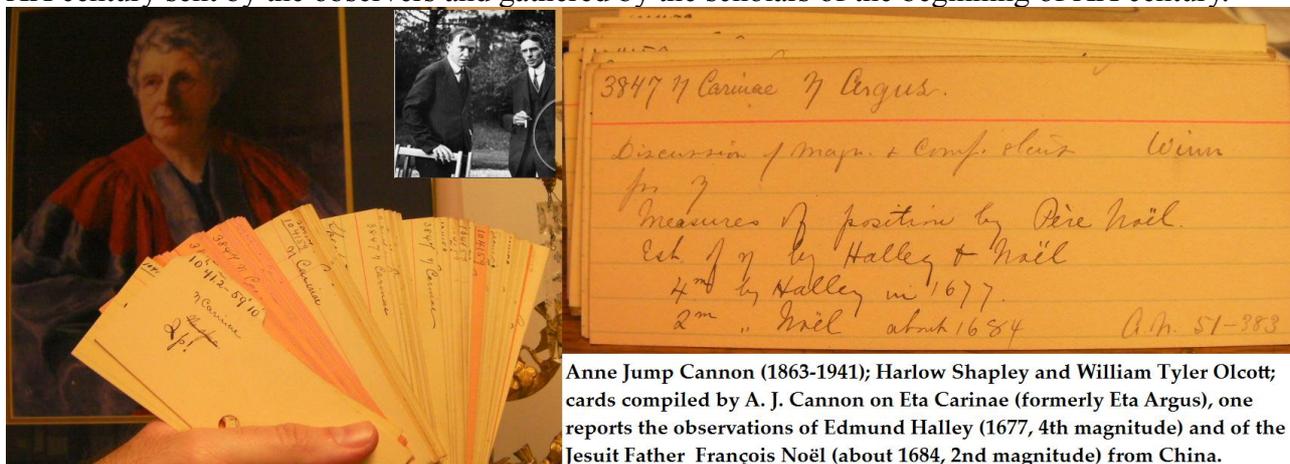

Anne Jump Cannon (1863-1941); Harlow Shapley and William Tyler Olcott; cards compiled by A. J. Cannon on Eta Carinae (formerly Eta Argus), one reports the observations of Edmund Halley (1677, 4th magnitude) and of the Jesuit Father François Noël (about 1684, 2nd magnitude) from China.

AAVSO slogan is "Citizen Astronomy": a modest equipment makes possible to contribute to science in astronomy with valuable data for future[5] that professional astronomers cannot gather.

---

1  T. R. Williams and M. Saladyga, Advancing Variable Star Astronomy, Cambridge University Press, UK, 2011.
2  ...it is a fact that only by the observation of variable stars can the amateur turn his modest equipment to practical use, and further to any great extent the pursuit of knowledge in its application to the noblest of the sciences.
   William Tyler Olcott, 1911.
3  Guthnick, P. 1901, Neue Untersuchungen über den veränderlichen Stern o (Mira) Ceti, Ehrhardt Karras, Halle a. S. , 241 pp.+25 graphs.
4  F. W. A. Argelander, The Variable Stars [in 3 parts], PA 20 (1912).
5  Dear Costantino,
   Congratulations on making and contributing your 1000th observation! That is truly an achievement. You say that there are other observers who are more productive - and yes, that is true - but we always say (and we believe) that quality is far more important that quantity. Each observation that is made well takes effort, care, and attention to

**Description of the first 1000 SGQ observations: analysis and accuracy**

**Mira:** The observations started in 1998 with Mira Ceti, that Dorrit Hoffleit (1907-2007) belonging to the first generation of American Variable Stars Observers and former director of M. Mitchell Observatory and Yale University Professor, called the "educational star". In 1999 in the occasion of my visit to Fermi Lab and Chile ESO I made some observations of Eta Carinae and of a Nova in Vela, and in the Centro Estudios Cientificos de Santiago, I applied for AAVSO membership, to send out these Southern Sky observations. Since then the observations of Mira continued to monitor 3 maxima, 1998-99; 1999-2000 and 2003 stimulated by my work on its variability history made in Yale with Dorrit Hoffleit and Riccardo Coccioli at Sapienza University. In this work the brightest maximum of Mira was the one observed by W. Herschel with Mira as brigth as Aldebaran, and our study wanted to find the probability of bright consecutive maxima, so an observational study of a few maxima paralleled the historical analysis. The database of Mira 1998-2003 from the personal astronomical diary was transferred online in June 2014; many of these observations are labeled as K, i.e. non-AAVSO charts, because I used other sources for comparison stars magnitudes. tot78 obs.

**Novae:** they are stimulating for their historical relevance and actuality of studies, as well as the Supernovae. Observing in Rome the limiting magnitude at the zenith is $Mv=4$ often I used binoculars and monoculars to limitate the influence of environmental city lights (direct) and to push of 1 or 2 magnitudes the limiting magnitude in case of faint stars ($Mv>3$). 8"SC telescope or 20x80 Binoculars have been used for stars $Mv>6$ as Mira at $Mv=9.2$ on 26/1 and at $Mv=8.87$-18/11/2014.

**Eta Carinae:** 37 obs from 2003 with some Eta Carinae observations from Rio, Copacabana (with strong light pollution and low security standard) to 2011 when I started in Rio, Flamengo, a long observational champaign on Delta Scorpii only one attempt to the Nova Puppis 2007 has been made, reserving all my observational time to the Sun and its diameter's variations. To Eta Car I made a new campaign from the Observatorio Nacional in Rio de Janeiro in 2013-4, measuring also V520 Car, of which only few observations from Hipparcos are recorded. The slow rising of luminosity of Eta Carinae has been noticed during the three observational campaign dedicated: June 1999 (4.9±0.1, July 2003 4.95±0.05 and December-April 2014: from magnitude 4.3 to 4.66).

From 2011 to 2015.5 I made 311 observations of **Delta Scorpii**, and since 2014 I follow Antares, to exploit the approach of Saturn, the sole object of similar magnitude within 45°: 160 obs to 2015.5

In opposite season and hours to Scorpius' appearance, I studied **Betelgeuse**, Alf Ori, another irregular variable of first magnitude, making 265 observations since 24/12/2011 in Paris, IAP.

While Betelgeuse is followed by hundreds fellows Antares is not, despite of the nomination as "star of the month" last June 2014; only my 100 observations are in the AAVSO database.

My naked eye observations of Alf Ori made from both hemispheres (Rio de Janeiro up to Warszawa, Poland) compared with digital observations in V band do not depart from them more than 0.05 magnitudes. Alf Ori is therefore the star with which I validate my personal techniques of observations which are mainly dependant on the choice of the reference stars.

For Alf Ori they are Procyon (Alf CMi) and Aldebaran (Alf Tau) within 20° and always corrected for airmass. For Antares they are much farther, up to 100°: Regulus (Alf Leo) and Spica (Alf Vir); only Saturn is now (2014-2015) closer to Antares, even if it is now 0.9 magnitude brighter.

**Extensions of temporal span of observations for Alf + Del Sco**

A problem for variable stars analysis is the gap when the Sun is near the star and the star is not visible. In the Fourier analysis this yields spurious periods. The attempts to extend the observational time of Delta Scorpii during the sunsets of september in Paris Institute of Astrophysics and october in Rome showed to me also the difficulty also for ancient peoples to follow the stars toward heliacal set and rise. The observation more intriguing in this respect has been Alf Sco on Nov 3, 2014 from Rome before sunset using binoculars aiming at the azimut of its setting and comparing with Altair

---

accuracy, and we value your efforts. Future researchers will also recognize your efforts and contributions as they work with your data perhaps many years from now. Thank you, Costantino!
best regards, Elizabeth O. Waagen, AAVSO June 9, 2015 -private communication.

(Alf Aql), and using magnitudes extinction calibrated with the Sun.[6]

The first day of January 2015 Antares was visible in the South Eastern horizon from Padua. This 2 months gap (Sun from 29°14' W to 30°54' E) is reduced to one month at the equator, where limiting the observations before/after the civil twilight is set (Sun at -6° below the horizon with the star up of 3-4°), Antares is not visible from Nov 17 to December 17.[7] Exploiting the SOHO-LASCO/C3 database, to find the stars in conjunction with the Sun in the ending days of November, we continued to observe Alf and Del Sco also from 23 Nov to 10 Dec.[8] 2013. Later SOHO LASCO C3 data have been used to sample Del Sco light curve from 1996 to 2013 during the solar conjunction. Summarizing the gap of Alf and Del Sco due to the solar conjunction has been reduced from about 60 full days to two periods of 20 days at 45° latitude, while this figure reduces at two gaps of 5 days (which are negligible in the Fourier analysis) for Equatorial observers, thanks to SOHO data.

**Novae and Supernovae: the comparison with instrumental data**

Finally we have three Novae monitored up to three months after the explosion: **Nova Del 2013**, 36 obs; **Nova Cen 2013** (with contributing observations from Sao Leopoldo, near Porto Alegre, Brazil, published in the IAUC of the discovery of that Nova) 40 obs. The multiple period star, R Cen, has been studied also during the Nova Cen campaign. The **Nova Sagittari 2015 no. 2**, currently in observation (7 obs) in June 2015 is dropping from 6.5 to 7.0 magnitude in an oscillanting phase.

The accuracy of the published observations -thanks to the gradual introduction of the airmass correction, extended from daylight measurements with the Sun (2003 and 2014), and to the study of the observations of the Novae made by very skilled observers, like Sebastian Otero, who gives the hundredth of magnitude in visual observations -is better than 0.05 magnitudes, despite of the environmental city lights experienced in all their variations from Rome to Rio de Janeiro, Observatorio Nacional and Copacabana, and the magnitude are evaluated always with two decimal digits since June/July 2014, while the 0.05 magnitudes quantization appeared one year before.

The agreement between SGQ observations and instrumental digital ones within 0.05 magnitudes shows the goodness of the procedure.

**Observational strategies and techniques**

Naked eye bright stars allow observations in every situation. The techniques have been reviewed by former AAVSO director Margaret W. Mayall in a manual for Variable Stars Observers.[9] Moreover they are stars easy to recognize to involve the largest number of students.

Beyond the standard technique of Argelander I have implemented in my observations the airmass correction in order to use stars very far from the target one, and at different altitudes. This permit to have comparison stars with a magnitude which changes continuously because of the atmospheric extinction. The computation of the magnitude of the variable star is made usually in conditions of equal apparent magnitude with the comparison star, which is rather easy to be judged with nake eye. The Stellarium software 12.4 and following versions with the extinction set to 0.23 magnitudes per airmass for Rome 60 meters above sea level, is used, with the recommendation to avoid stars lower than 15° degrees, where the computation need to be checked. To find Novae the AAVSO Charts, with the Mappa Stellare for Ipad (with Hipparcos star in) have been used together with significant speed up of the pointing. Telescopes in fixed position are naturally better for pointing faint objects. The software Ephemvga has been used for precession calculations, Julian Date and ephemerides.

---

6  Thanks to prof. M. Climaco for the technical assistance and 2CT of G. Ferraris institute of Rome for the experiment in Villa Pamphili, on October 27 2014 from 14 to 17h.
7  Simulations made with Stellarium 12.4; Saturn visual magnitudes are better in http://www.curtrenz.com/saturn.html
8  Alexandre Amorim, Observation Coordinator of Núcleo de Estudo e Observação Astronômica "José Brazilício de Souza" and AAVSO member of Santa Catarina State, and I, during the LARIM 14 (Latino American Regional IAU Meeting, started in the feast of St. Catarina of Alexandria, patron of the State on 25/11) in Florianópolis, Brazil, we observed the comet C/2012 S1 ISON grazing the Sun on November 28, 2013. We followed it on SOHO images and with binoculars on tripod at the congress site and at Sacred Hearth Parish of Ingleses, Florianopolis, evaluating correctly its magnitude Mv>-2 and its disappearing after the graze. Amorim wrote a tutorial for naked eye observers.
9  It is available in pdf at the site http://www.as.up.krakow.pl/gzz/teksty/manual.pdf Manual of Visual Observing of Variable Stars (revised edition 2001)

**Novae and Supernovae: physical basis for interpreting the data**

The ejecta travel at non relativistic speeds v~c/300, but the time compression effects of Beyond the debate of the standard candle nature of SN1a for measuring the accelerating Universe,[10] the exponential decay is a general law in physics applied to stellar explosions and it is evident in our database of three Novae observed along 3 months after the explosion and a Supernova.

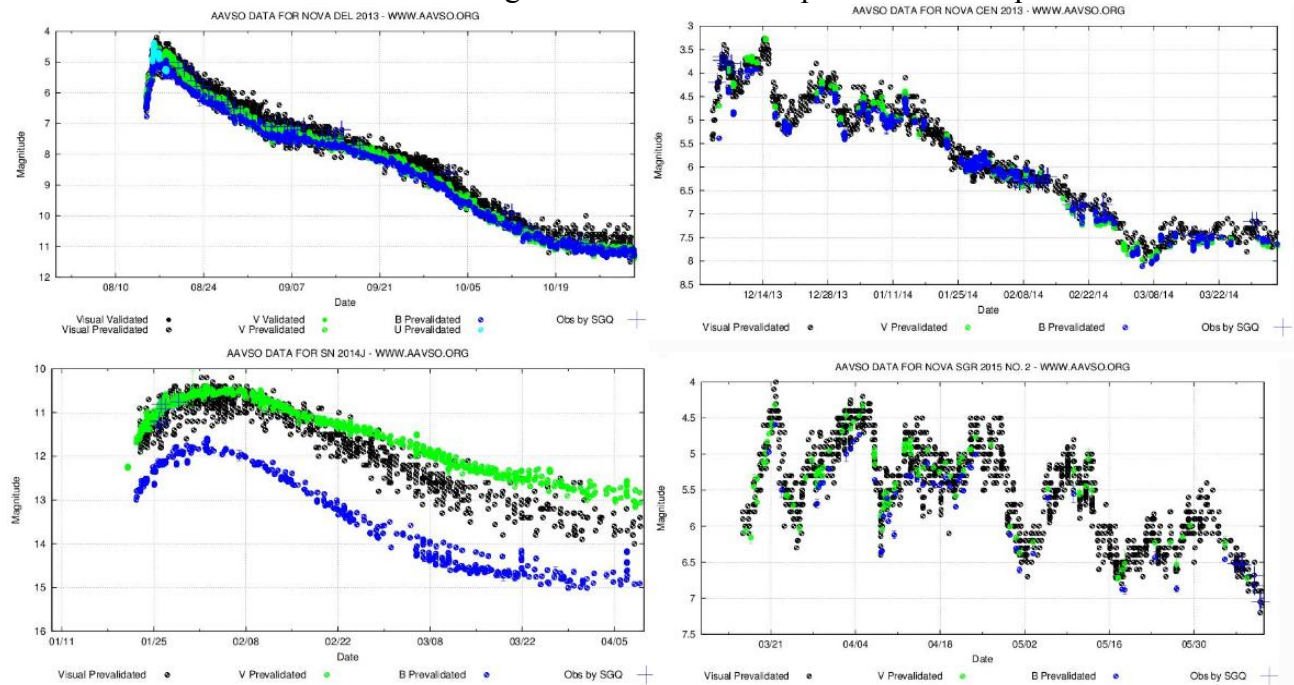

On the left Nova Del 2013 up and **SN2014J** down, on right side Nova Cen 2013 and down Nova SGR 2015 with more irregular descent. With my naked eye of December 4th 2013 observations I coauthored the IAUCircular of the discovery of Nova Cen 2013. A table with the first 8 stars follows

| Mira Ceti | Del Sco | Alf Ori | Alf Sco | Nova Del | Nova Cen | SN2014J | Eta Car |
|---|---|---|---|---|---|---|---|
| 1998-78 | 2011-311 | 2011-265 | 2014-160 | 2013-36 | 2013-40 | 2014-8 | 1999-37 |
| Roma | Rio de Jan | Paris | Atlantic ocean | Roma | S.Leopoldo | Roma | Santiago |

The location of the first observation has been included in the table. Alf Sco was first measured onboard AF442 flight above Atlantic Ocean. Other stars followed are R Cen 22; V520 Car 18; V766Cen 9; Del Cep 12 and Alf Com 3 during the 2015 eclipse campaign and Nova SGR 2015#2 7. The notes for each observation are useful for recomputing airmass correction and characterizing an observing site in terms of useful nights vs not (especially for Rome, Paris and Rio de Janeiro).

**References**

Amorim, A.   Boletim Observe! Janeiro 2014.on the Comet C/2012 S1 ISON
Argelander F. W. A. 1869, Astronomische Beobachtungen auf der Sternwarte, VII, A.Marcus, Bonn.
Guthnick,P. 1901, Neue Untersuchungen über den veränderlichen Stern o(Mira) Ceti,E.Karras,Halle
Izzo, L. et al, Spectroscopic observations of Nova Cen 2013, ATel 5639 (2013).
Mayall, M.,AAVSO Manual for Visual Observing of Variable Stars, AAVSO 2001.
Moore, Patrick, Naked eye Astronomy, Luttherword Press, London, 1976.

Sigismondi, C., R. Coccioli and D. Hoffleit JAAVSO, 30, 31 (2001)
Sigismondi, C., in Quodilibet: http://www.quodlibet.net/articles/sigismondi-mira.shtml  (2002)
Sigismondi C., et al., arXiv 1410.8492 Del Sco with SOHO LASCO/C3
Sigismondi, C., Considerations on the Nova Delphini Light Curve, arXiv 1310.2763S
Guido, E., et al., IAUC 9265, V1369 Centauri = Nova Cen 2013 (2013)
Sigismondi, C., arXiv 1312.4848 on Nova Centauri 2013 broad maximum
Thomas, H. L. 1948, The Early History of Variable Stars Observing, PhD Thesis, Radcliffe College.

---

10  http://www.ptf.caltech.edu/news/14